\documentclass[preprint,12pt]{elsarticle}

\usepackage{amssymb}
\usepackage{amsthm}

\usepackage{lineno}

\usepackage[T1,T2A]{fontenc}
\usepackage[polish, english]{babel}
\usepackage{hyperref}
\usepackage{url}
\usepackage{xurl}

\journal{Chaos, Solitons \& Fractals}

\begin{document}

\begin{frontmatter}



\title{Shape patterns in popularity series of video games}


\affiliation[inst1]{organization={Departamento de Física},
            addressline={Universidade Estadual de Maringá}, 
            city={Maringá},
            postcode={87020-900}, 
            state={PR},
            country={Brazil}}

\author[inst1]{Leonardo R. Cunha}
\ead{leoribeirocunha@gmail.com}

\author[inst1]{Arthur A. B. Pessa}
\ead{arthur\_pessa@hotmail.com}
\author[inst1]{Renio S. Mendes}


\begin{abstract}
In recent years, digital games have become increasingly present in people's lives both as a leisure activity or in gamified activities of everyday life.
Despite this growing presence, large-scale, data-driven analyses of video games remain a small fraction of the related literature.
In this sense, the present work constitutes an investigation of patterns in popularity series of video games based on monthly popularity series, spanning eleven years, for close to six thousand games listed on the online platform Steam. 
Utilizing these series, after a preprocessing stage, we perform a clustering task in order to group the series solely based on their shape.
Our results indicate the existence of five clusters of shape patterns named decreasing, hilly, increasing, valley, and bursty, with approximately half of the games showing a decreasing popularity pattern, $20{.}7\%$ being hilly, $11{.}8\%$ increasing, $11{.}0\%$ bursty, and $9{.}1\%$ valley.
Finally, we have probed the prevalence and persistence of shape patterns by comparing the shapes of longer popularity series during their early stages and after completion.
We have found the majority of games tend to maintain their pattern over time, except for a constant pattern that appears early in popularity series only to later originate hilly and bursty popularity series.
\end{abstract}




\end{frontmatter}


\section{Introduction}\label{sec:introduction}

Social groups and communities are largely understood through cultural content. 
In general, cultural content refers to any form of information, expression, and creativity that reflects or represents elements of a particular culture~\cite{turp2013contribution}. Cultural content can encompass literature, art, music, film, television, dance, folklore, traditions, rituals, language, games and so on. 
It also plays a crucial role in shaping and maintaining cultural identity, fostering a sense of community, and transmitting knowledge from one generation to the next~\cite{ferdman1990literacy, fernandez2018rediscovering}. 
Moreover, as societies evolve and interact, cultural content can also be influenced by and influence other cultures, leading to a dynamic and interconnected global cultural landscape~\cite{griswold2012cultures}. Especially in recent years, driven by technologies and digital tools, the dissemination of cultural content has become even more intense and accelerated~\cite{lorenz2019accelerating}. 
In extreme cases, phenomena have emerged such as best-sellers~\cite{sorensen2007bestseller}, viralization~\cite{berger2012makes}, memes~\cite{gleeson2016effects, d2019spreading}, and fake news~\cite{lazer2018science, vosoughi2018spread}. 
In this way, while cultural content aids in characterizing a society, the dissemination of such cultural content can affect the collective behavior of that same society~\cite{paek2004spreading}.

Although the digital era is a recent development, the digital and automated storage of data establishes a conducive environment for the advancement of research on patterns of our collective behavior~\cite{salganik2019bit}. Anchored in large databases, several works use ``measures of consumption'' as proxies for the amount of collective attention that a type of cultural content receives. 
As examples, we have the number of shares of hashtags on Twitter~\cite{chang2010new, bruns2013towards}, movie ticket sales~\cite{lorenz2019accelerating}, citations in scientific publications~\cite{eom2011characterizing}, music streams on Spotify~\cite{bello2021cultural}, views of videos on YouTube~\cite{hoiles2017engagement}, and the number of page views on Wikipedia~\cite{miz2017wikipedia, moat2013quantifying}. Thus, the idea of collective attention is closely related to the notion of a computational social science that investigates various patterns of content consumption at a population level~\cite{crane2008robust, eom2011characterizing}.

In this context, video games are perhaps the principal cultural product of our digital era with far-reaching implications for culture, most especially, through the gamification of several aspects of modern life, from work to learning~\cite{muriel2018video}.
In addition, what in the past was seen as child's play is now a widespread pastime among young people and adults, regardless of gender~\cite{jovanovic2022gamer}, making digital games an increasingly popular form of entertainment and leisure activity.
Harnessing the power of entertainment, the gaming industry has proved to be very profitable over the last few years, bringing in billions of dollars~\cite{newzoo2021}.
Despite this cultural and economic relevance, most works related to video games frequently utilize small samples, are qualitative, or focus on certain characteristics of gamers and games~\cite{muriel2018video}.
For instance, considering works developed from controlled experiments, researchers have measured people's motivation throughout the stages of a game~\cite{strojny2023player}. 
Using self-reports, studies determined factors related to the gain or loss of individual motivation to continue playing~\cite{jin2012toward, kahn2015trojan}.

In a more quantitative vein, there exists a more recent, emerging literature related to video games based on statistical analyses of large amounts of data.
These works have, for example, investigated behavioral characteristics of players across different games and countries~\cite{sifa2015large-scale, zendle2023cross, cunha2024complexity}, studied textual characteristics of game reviews~\cite{lin2018empirical, lin2019empirical}, and looked into time series of game popularity~\cite{ahn2017makes, mendes2022popularity}.   
Knowing that the objective of game developers is to get people to play their games, here, we consider the time series of average hourly number of players of a game as a proxy for its popularity, a proxy for the collective attention received by it, in order to detect groups of games that display similar popularity patterns.
To do so, we have adopted a clustering approach, inspired by recent works that apply clustering algorithms to investigate diverse characteristics of time series of complex systems related to cryptocurrencies~\cite{gidea2020topological, sigaki2019clustering}, scientific careers~\cite{sunahara2023universal}, and dynamical systems~\cite{pessa2022clustering, zhou2022recognition}.
Furthermore, we have utilized the Steam's database, which provides the average hourly number of players per month for 5,840 games in order to detect groups of games whose player count time series display a similar shape pattern of temporal evolution.
We highlight the novelty of our work in the scientific literature concerning digital games, given our approach and magnitude of dataset.

The remainder of this paper is organized in another four sections.
We proceed, in Section~\ref{sec:dataset}, by describing our database of popularity series of video games. We also model the growth in the cumulative number of released games and their popularity and fit the distribution of average popularities for all games.
In Section~\ref{sec:methods}, we detail our statistical approach to group these popularity series according to their shape and present the obtained results in Section~\ref{sec:results}. 
In Section~\ref{sec:discussion}, we discuss the implications of these results, the consistency of our findings with the literature, suggest avenues for future work, and summarize the main results before concluding.

\section{Dataset}\label{sec:dataset}

\begin{figure*}[!ht]
\centering
\includegraphics[width=1\linewidth]{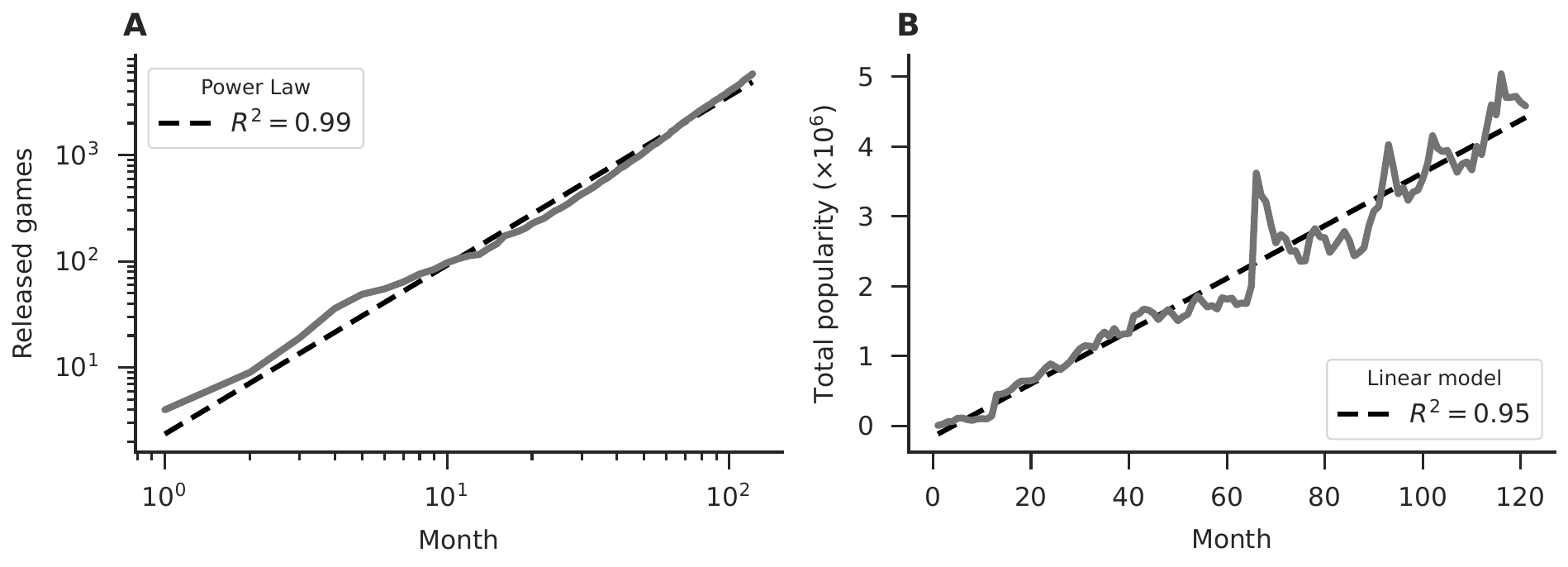}
\caption{{\bf Temporal evolution of the number of games and total popularity.} (\textbf{A}) The cumulative number of games monthly released on Steam is represented by a solid line and the dashed line stands for a power-law fit with a power-law exponent equal to $1{.}81$. (\textbf{B}) The sum of popularities of games for each month is represented by a solid line and the dashed line stands for a linear fit. In both panels, the data covers a period of 120 months from August 2012 to July 2022.}
\label{fig:1}
\end{figure*}

Our results are based on Steam game data. Steam is an online platform which sells and manages the copyrights of digital products from its base~\cite{steam2020} and has millions of users spread all over the world~\cite{steam2019stats}, as is perhaps the principal source of the data utilized in more quantitative investigations involving games and gamers~\cite{lin2018empirical, lin2019empirical, sifa2015large-scale}.
Since the Steam service is completely online, it can store an enormous amount of data about its users in an automated way. In this sense, some third-party websites make part of this data openly available. To obtain our results, we use the data provided by the website SteamCharts~\cite{gray2016steam}.
SteamCharts is a website launched in July 2012 that publishes the average hourly player count for each month of Steam games.
For example, if a game has a popularity value equal to 100 in a given month, this means that, on average, 100 people were playing the game each hour during that month.
Combining Steam and SteamCharts, we get the name, release date, genre, and popularity time series of games between August 2012 and July 2023. 
Filtering out games that were missing some information, games that were released before August 2012, and games with less than 1 year of time series data (games released after July 2022), $5{,}840$ games have remained to be studied. 
The most important information of each game is the time series of the average hourly number of players per month, which we call its popularity series, or simply its popularity.

We begin exploring our dataset with two of the main variables concerning Steam: the number of games and the total popularity.
Figure~\ref{fig:1}A shows the cumulative number of games released from August 2012 to July 2022. As can be seen, the total number of released games exhibits a superlinear growth, which can be fitted by a power law with very good approximation.
Figure~\ref{fig:1}B depicts the temporal evolution of the sum of monthly popularities of all games on Steam.
This metric reveals an approximately linear growth in the total popularity of Steam games over time.
An interesting aspect of the Steam popularity dynamics is the fluctuation, especially in the second half of the curve, where many distinct peaks emerge prominently.
In particular, the most distinct peak occurs in month 66 (January 2018), 
reaching a total popularity of 3,621,887. Taking into consideration the time span covered by our data, the overall maximum occurs in March 2022, reaching a total of 5,043,435. 
Overall, the results from Fig.~\ref{fig:1} highlight the growth of Steam in recent years and of gaming as a hobby or leisure activity.

\begin{figure*}[!ht]
\centering
\includegraphics[width=1\linewidth]{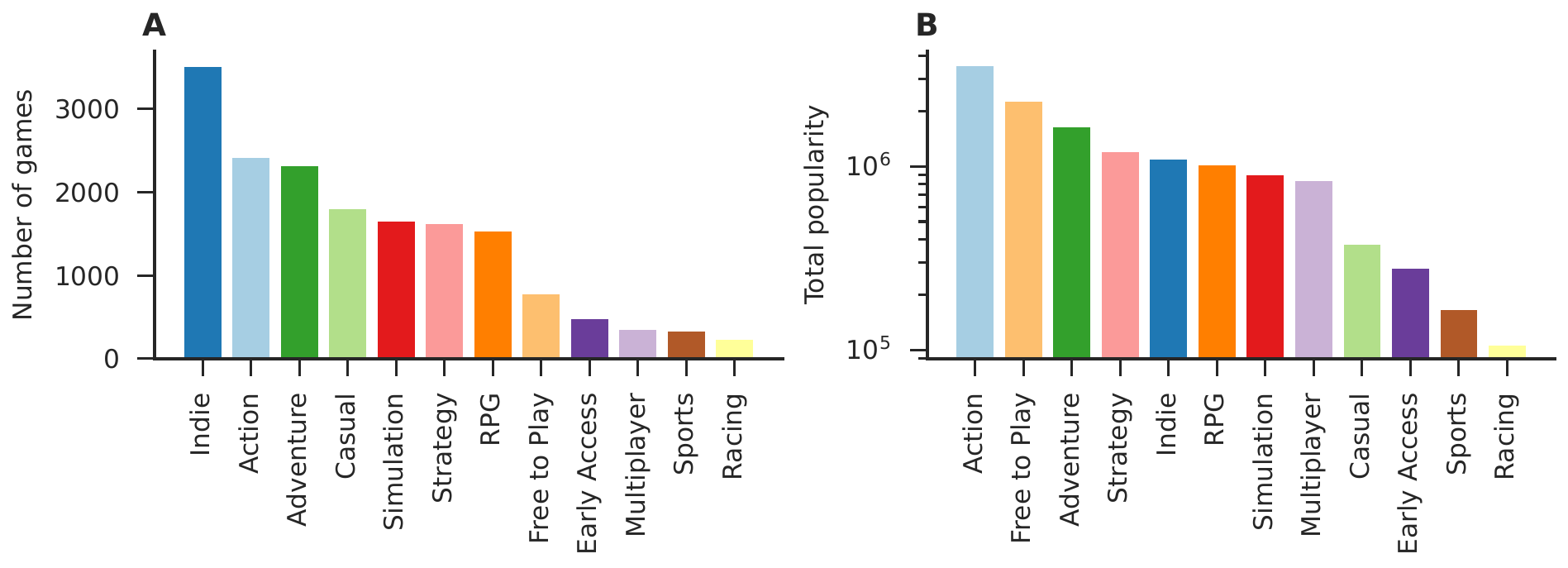}
\caption{{\bf Number of games and players related to each game genre.} (\textbf{A}) Cumulative number of games released on Steam belonging to each genre until July 2022. (\textbf{B}) Number of players per genre in July 2022. For each genre, the popularity of games in the month of July 2022 was summed.}
\label{fig:2}
\end{figure*}

Investigating the same variables on a different scale, we have analyzed the number of games for each genre on Steam. When releasing a game, its developer classifies it within some genres, similarly to what happens with movies. There are 12 official genres on Steam: Action, Adventure, Casual, Early Access, Free to Play, Indie, Massively Multiplayer (abbreviated as Multiplayer), Racing, Role Playing Game (abbreviated as RPG), Simulation, Sports, and Strategy. Among these 12 genres, developers choose one or more to characterize their games. 
In Fig.~\ref{fig:2}A, we show the amount of games released until July 2022 in each genre. The genre with the largest number of games is Indie, which does not characterize a style, but indicates games independently produced, making it the genre with the highest number of games released, since they are produced by small companies and, in general, are simpler games. In this line of genres that do not define a style, there are also other genres such as Free to Play (free games) and Early Access (games available to play in their beta phase, preceding the official launch). Among genres that characterize aspects of immersion, broader genres such as Action and Adventure represent the majority of games. On the other hand, the smallest number of games released belong to the Racing genre, probably because this category is much more restricted and defined. Taking into account temporal evolution, all genres have grown with superlinear trends too, and a large portion of the genres can be well approximated by a power-law model (see Fig.~S1).

Deepening the genre investigation, we have looked into the number of players per genre to get an intuition of people's preferences. In Fig~\ref{fig:2}B, we show the aggregated popularity of games for each genre in the month of July 2022. It's worth noting that a game with more than one genre contributes to the popularity of all the genres it falls under. In spite of that, we've noticed that players have a strong preference for Action games, which is possibly the reason why developers make a lot of games of this genre. Consistently with what Indie stands for, despite having many games released, its popularity relative to other genres is middling. As already mentioned, independent productions usually mean simpler games that end up not reaching the attention of players. On the other hand, there are far fewer Free to Play games and they reach a much wider audience, being the second most popular genre. Even though Action has more players, Free to Play games achieve a very high engagement rate, because in raw numbers, Action has 1,354 games and in July 2022 it had 3,970,486 total popularity, while Free to Play had 580 games and 2,504,271 total popularity during the same month. It seems reasonable to think that free games have greater reach and therefore greater popularity. At the other extreme, the least popular genre is Racing. Analogously, we can use the same argument used in the previous paragraph about genre specificity. Examining in more detail, Fig.~S2 shows the temporal evolution of the number of players in each genre.
Despite all genres showing increasing popularity, the growth characteristics are slightly different for each one of them, with very pronounced local peaks standing out (as in Figure~\ref{fig:1}{B}). 
Focusing on the most significant peak, which occurs in the 66th month (January 2018), we observe that it is the result of the official release of a game, named PUBG. 
This game (belonging to the genres Action, Adventure, Free to Play, and Multiplayer) achieved the highest popularity in a single month throughout the entire database, and then its popularity declined quite rapidly over the following months (see Fig.~S3).

\begin{figure*}[!ht]
\centering
\includegraphics[width=1\linewidth]{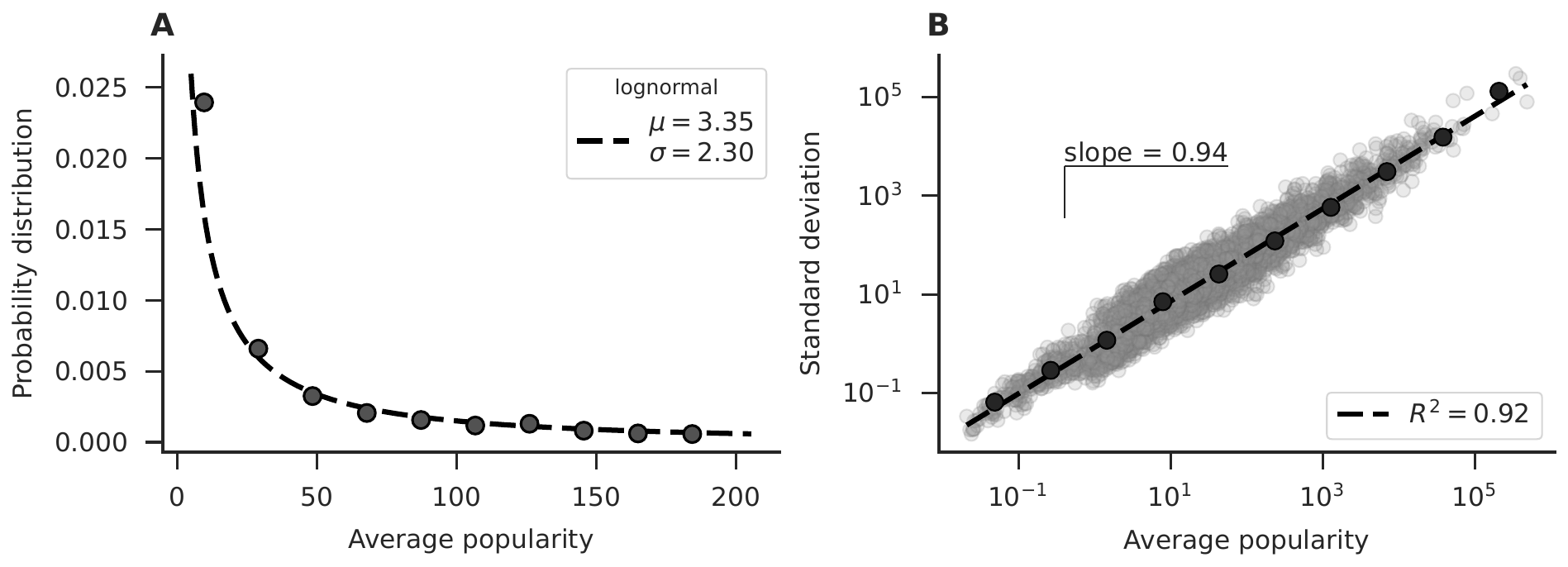}
\caption{{\bf Summary of time series of game popularity.} (\textbf{A}) Probability distribution of the number of games as a function of the average of the popularity series. The gray circles represent the data and the dashed line stands for a log-normal distribution of parameters $\mu = 3{.}35$  and $\sigma = 2{.}30$. (\textbf{B}) Relationship between the averages and standard deviations of popularity series. The light gray circles are the data, the black circles are averages taken in equally spaced logarithmic windows, and the black dashed line represents a linear model fitted to the data ($R^2 = $ 0.92).}
\label{fig:3}
\end{figure*}

A simple inspection of popularity series shows that most games have few players, while a few games concentrate a huge mass of players. Looking more closely, we find that approximately 73\% of popularity series have an average of fewer than 100 active players per hour per month. 
On the other hand, the game with the highest average popularity, Dota 2, has had around 485,000 active players per hour throughout its history. 
Fig.~\ref{fig:3}A exhibits the distribution of the number of games as a function of the average number of players, where the distribution presents a heavy tail. 
The dashed line represents a log-normal distribution with parameters $\mu = $ 3.35 and $\sigma = $ 2.30, the best fit to data according to the Kolmogorov-Smirnov statistical test~\cite{massey1951kolmogorov} (see Table~S1 for more details and other distributions tested).
The average is perhaps the most common summary statistic calculated from a random variable, and it is usually paired with a measure of dispersion such as the standard deviation. 
In this direction, we also calculate the standard deviation for each popularity series and show its relationship with the mean. 
In Fig.~\ref{fig:3}B, we see, as a general trend, that the more the average popularity increases, the more the standard deviation increases.
Interestingly, according to our fit (represented by the black dashed line), the growth is approximately linear, as the slope is 0.94 $\approx$ 1 on the log-log scale, possibly suggesting an anomalous process of ``attention diffusion''~\cite{wagenmakers2005relation}.

\section{Methods}\label{sec:methods}

In a next step, we characterize shape patterns in time series of popularity of video games, that is, the way a popularity series evolves over time, without taking into account its magnitude, only its shape.
In this way, we group series that present similar shapes, even if they do present average hourly player counts on different scales. 
Targeting this shape-based clustering approach, we transform each popularity series into a $z$-score. 
Thus, we use series whose variation is given in standard deviation units in relation to the mean of the series itself. 
However, the $z$-score series is very noisy, showing a large variation in the number of players from one month to another. Because of this, we apply a smoothing kernel to the series, called a Gaussian filter~\cite{ito2000gaussian}, to reduce random fluctuations and make the pattern of the series more evident. The Gaussian filter depends on the parameter $\sigma$, and this parameter is related to the degree of smoothing. The larger the value of $\sigma$, the smoother the series becomes. In the supplementary material, we show examples of popularity series alongside their $z$-score series and Gaussian filters with $\sigma = $ 1, 2, and 3 (see Fig.~S4). For $\sigma = $ 1, some series still remain quite noisy; for $\sigma = $ 2, the series become smooth without losing their shape, whereas for $\sigma = $ 3, the series are so smooth that they lose the amplitude of variation and some important details of the shape. 
Taking these observations into consideration, we have chosen $\sigma = $ 2. 

From these smoothed $z$-score series, we calculate the distance matrix between them using the Dynamic Time Warping (DTW) algorithm~\cite{sakoe1978dynamic, meert_wannes_2020_7158824}. 
Generally, the DTW distance is a shape-based dissimilarity measure that allows us to compare series of different sizes by creating a non-linear warping path between them, providing more flexibility for matching sequences that show similar patterns.
After calculating the distance matrix, we apply the Uniform Manifold Approximation and Projection (UMAP) method~\cite{mcinnes2018umap, mcinnes2018umap-software} (with default parameters) to perform a dimensionality reduction.
In a nutshell, from the dissimilarity matrix, UMAP performs a dimensionality reduction based on mathematical foundations of Riemannian geometry and algebraic topology. 
UMAP first creates a fuzzy simplicial complex, which can be represented as a weighted graph, and then projects the data into a lower-dimensional space via a force-directed graph layout algorithm. 
Thus, the algorithm creates a network representation of the dissimilarity matrix, where nodes represent games and weighted edges connect games with similar dynamic popularity.
This dimensionality reduction technique is designed in the form of a complex network, arranging and linking nodes in order to balance the arrangement respecting the local and global structures. 
Here, however, we dispense with the graph associated with the low-dimensional space and apply the Infomap algorithm to identify the community structure of the first network created by UMAP. 
Infomap is a network clustering technique based on concepts of information theory that relies on random walks as a proxy for information flow over the network. 
Within the algorithm is calculated a map equation~\cite{rosvall2008maps, rosvall2009map, Edler_The_MapEquation_software_2022}, the theoretical limits of how concisely one can describe an infinite random walk on the network (the description length) with a particular assignment of nodes to communities. 
Minimizing the map equation, Infomap uncovers the network's community structure. 
As the Infomap algorithm is based on the probability flow of random walks on the network, different runs lead to the detection of different community structures. 
In order to escape this problem, we run ten thousand realizations of the Infomap algorithm.
In these runs, we have detected numbers of clusters ranging from $3$ to $7$, with $5$ being the most recurrent number of clusters detected (for more details see Fig.~S5). 
Among these ten thousand runs, we have chosen the one that expresses the best cluster partition based on the silhouette score~\cite{rousseeuw1987silhouettes} calculated with the values on the DTW distance matrix. 
This coefficient quantifies the consistency of the clustering procedure, giving a measure of the similarity of the elements of the same cluster and the dissimilarity of the elements of different clusters. 
Generally speaking, the higher the average value of the silhouette score, the better the cluster configuration.

Therefore, after all these steps, we can cluster popularity series with similar shapes. 
This same algorithmic routine was recently proposed by Sunahara \textit{et al.}~\cite{sunahara2023universal} as a modification of the algorithm of Lee \textit{et al.}~\cite{lee2021non}. 

\begin{figure*}[!ht]
\centering
\includegraphics[width=1\linewidth]{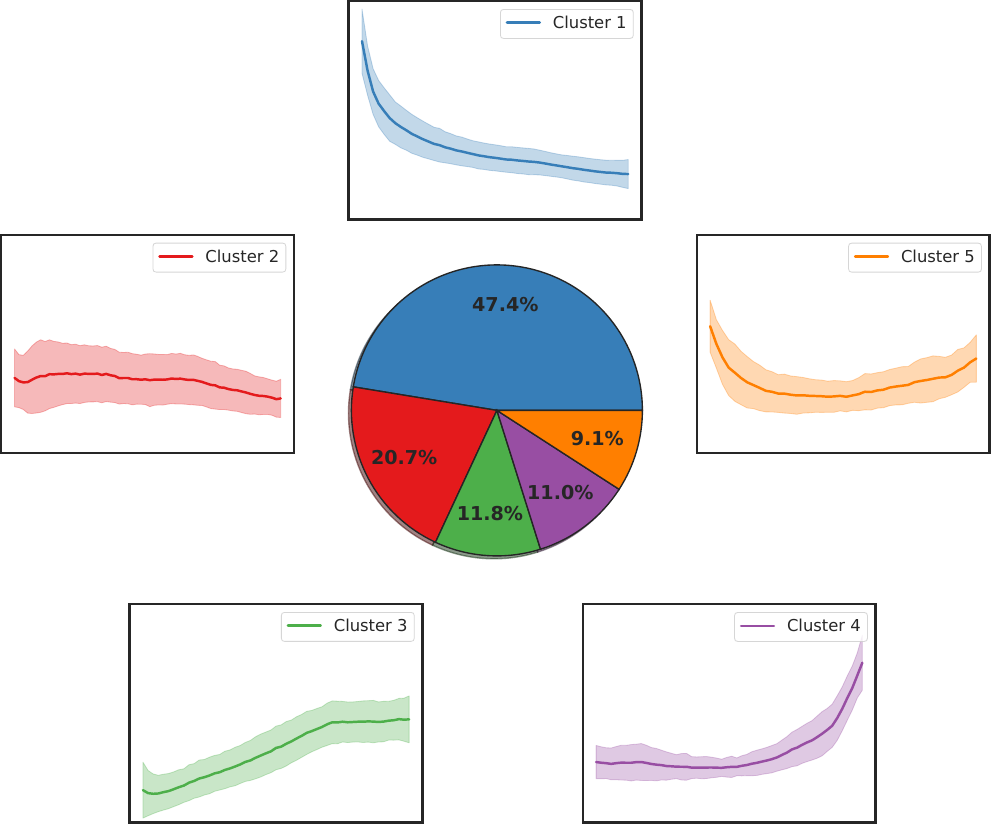}
\caption{{\bf Shape patterns of game popularity series on Steam.} The center panel shows the proportion of games belonging to each group that was detected by the clustering algorithm: 47.4\% of games are in cluster 1 (blue); 20.7\% in cluster 2 (red); 11.8\% in cluster 3 (green); 11.0\% in cluster 4 (purple); 9.1\% in cluster 5 (orange). Panels around the center plot depict the average shape of times series belonging to each cluster and one standard deviation band.}
\label{fig:4}
\end{figure*}

\section{Results}\label{sec:results}

The best clustering of popularity series and the average shape of series pertaining to the clusters we have found are shown in Fig.~\ref{fig:4}.
In total, we have detected 5 clusters, each representing a different evolution of popularity. 
Cluster 1 (blue), which accounts for 47.4\% of the games, displays a predominant pattern of decay, making it clear that a significant portion of games reaches its peak popularity in the first few days after launch and then gradually fades over time. 
Cluster 2 (red), representing 20.7\% of games, exhibits the most irregular pattern among all clusters. 
The most distinctive feature of this cluster is its hilly pattern, showing multiple peaks of popularity over time. In Fig.~S6, we show two typical popularity series for each cluster. Despite the two series illustrated as examples of cluster 2 appearing to have a declining trend, some series have shown an approximately constant trend, and in other cases, there may even be a slight growth. Thus, we observe that the most distinctive feature of the series in cluster 2 is its multiplicity of popularity peaks that occur from time to time. 
Cluster 3 (green), 11.8\% of the games, exhibits a pattern of consistent and enduring growth. In addition to having a growth pattern, games in cluster 3 boast the highest average of popularity (see Fig.~S7). Nevertheless, it is worth noting that in a large number of games in this cluster, the initial growth is more accelerated compared to the last third of the time series.
In fact, for many games, there is a saturation in the last months (see examples in Fig.~S6). 
Cluster 4 (purple), 11.0\% of the games, consists of stable and relatively unknown games for the most part; however, from a certain month onwards, they display a burst in popularity. 
Cluster 5 (orange), 9.1\% of the games, displays the most unexpected pattern, initially declining and then followed by a recovery in popularity. Due to this popularity pattern resembling a valley, we hypothesize that games in this cluster would have longer series, as it would allow these games to be forgotten and then remembered as ``good games''. However, Fig.~S8 shows the opposite; cluster 5 exhibits the greatest concentration of probability density at lower values of series lengths, with the smallest median among all clusters.

\begin{figure*}[!ht]
\centering
\includegraphics[width=0.82\linewidth]{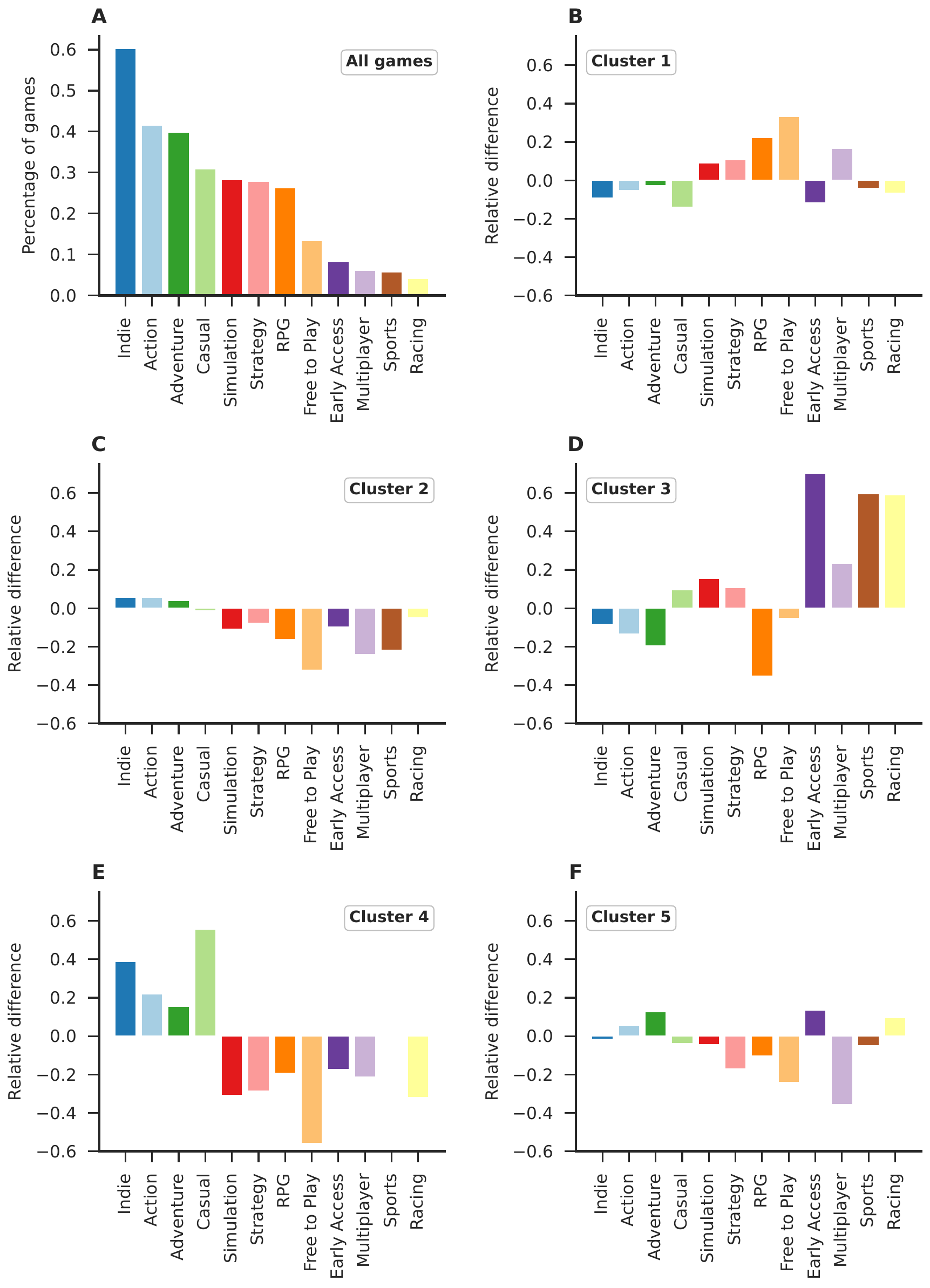}
\caption{{\bf Representation of genres and their relative proportion in the detected clusters.} (\textbf{A}) Percentage of games in our database belonging to one or more of the 12 official genres on Steam. Relative differences in representation of game genres for games pertaining to shape clusters 1 to 5 are shown in panels (\textbf{B})-(\textbf{F}).
}
\label{fig:5}
\end{figure*}

The representation of game genres varies across clusters. 
To examine this aspect, we first calculate the difference between the proportion of games for a given genre within each cluster and the overall proportion of games of the same genre (Fig.~\ref{fig:5}).
Furthermore, we divide this difference by the overall proportion of games belonging to the genre.
Fig.~\ref{fig:5} illustrates this difference in genre representation within each cluster. Indie and Casual games are overrepresented in cluster 4, suggesting that a portion of these games achieves ``success'' later on. On the other hand, Simulation and Strategy are underrepresented in cluster 4, suggesting that their popularity patterns are more consistent over time, and therefore, it is rarer for sleeping beauty phenomena to occur in these genres. Meanwhile, RPG seems to be a game genre prone to being completely forgotten; it is overrepresented in cluster 1 (decreasing) and underrepresented in the other clusters, especially in cluster 3 (increasing). Interestingly, Free to Play games are overrepresented in cluster 1 and underrepresented in the others as well; however, they are very popular games (Fig.~\ref{fig:2}). 
This suggests that, in many cases, these games have a meteoric start and then gradually fade. The most surprising finding was the remarkable representation of Early Access games in cluster 3. Multiplayer games exhibit stable dynamics of popularity growth or decline, as these games are overrepresented in clusters 1 and 3, and underrepresented in the others. The two more niche genres, Sports and Racing, are overrepresented in cluster 3, demonstrating the potential of their attractive force in capturing new players. The two genres that did not exhibit notable overrepresentation nor underrepresentation in any cluster were Action and Adventure.

\begin{figure*}[!ht]
\centering
\includegraphics[width=0.85\linewidth]{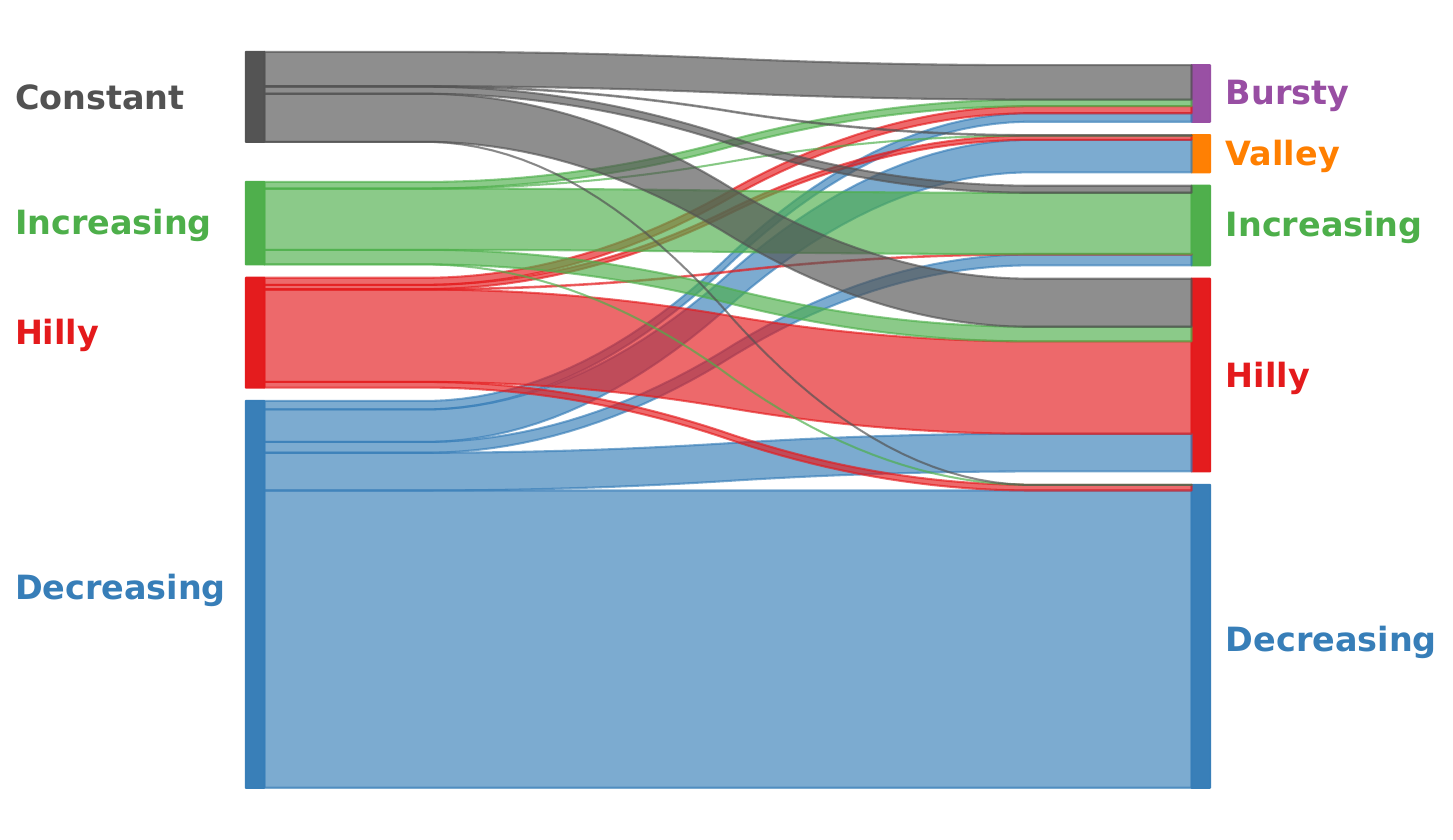}
\caption{{\bf Comparison between the prevalence and persistence of shape patterns in the early and final stages of popularity series.}
Left bars show the fraction of games belonging to each cluster of shape pattern in the first 24 months, while the right bars show the prevalence of these shape patterns considering the complete popularity series. 
Connections between patterns on the left and right signify the migration from one pattern to another as popularity series evolve over time. Additionally, the thickness of the links represents the proportion of games.}
\label{fig:sankey_plot}
\end{figure*}

In the next step, we analyze whether popularity patterns are related to series length. 
To do this, we divide the popularity series into four groups of same size based on quartiles from the distribution of lengths (as shown in Fig.~S9).
The four groups encompass games whose series are between 12 to 27 months long, 28 to 49 months, 50 to 74 months, and games with series longer than 74 months.
Fig.~S10 shows how each length group is composed of games belonging to different popularity clusters. 
We observe the percentages of games with decreasing (cluster 1) and valley (cluster 5) popularity patterns decrease as their series become longer.
However, the decreasing pattern remains the most recurrent, independently of series length.
The hilly pattern (cluster 2) sees an increase in representation as series become longer. 
Surprisingly, the bursty pattern (cluster 4) shows a significantly higher percentage of series between 50 to 74 months long.
Meanwhile, the increasing pattern (cluster 3) maintains its percentage across different lengths, suggesting that ``successful games'' are released at a fixed rate.

In the final part of our analysis, we investigate the prevalence and persistence of popularity patterns over time. 
In other words, for all games, we compare the popularity pattern in the initial stages to that of the whole series. 
Considering the group of older series (longer than 75 months), we employ the aforementioned cluster detection method (Sec.~\ref{sec:methods}) to identify the popularity patterns of the initial 24 months. 
Fig.~S11 shows the frequency of detected clusters after ten thousand realizations of the Infomap algorithm. Meanwhile, Fig.~S12 displays the detected clusters using the best realization according to the silhouette score. In these detected clusters, three patterns have already been exhibited: decreasing (cluster 11 and cluster 14), hilly (cluster 12), and increasing (cluster 15). 
The novelty was the emergence of a constant shape pattern with little variance (cluster 13). 
The valley and bursty patterns were not detected. Connecting the results of the cluster analysis for the initial stages with the findings previously presented for the whole series, we measured the persistence of patterns. 
Restricting our analysis only to games older than 75 months, Fig.~\ref{fig:sankey_plot} shows the popularity patterns exhibited by games considering the first 24 months (left bars) and the whole series (right bars). 
We find that 77\% of games maintain the decreasing pattern, 84\% remain hilly, and 74\% remain increasing.
The most significant changes occur in games that exhibit constant and decreasing shape patterns at the beginning of their popularity series. 
Games with constant pattern basically become hilly (53\%) and bursty (37\%), while 10\% of the decreasing ones become hilly and 8\% become valley. Examples of popularity series of games undergoing pattern changes over time can be seen in Fig.~S13.

\section{Discussion and Conclusions}\label{sec:discussion}

We have presented a quantitative study on digital games and their players. Specifically, we have employed a cluster analysis approach to detect patterns in the dynamics of game popularity. To do this, we have used the average hourly number of players per month for 5,840 games on Steam. 
In our exploratory analysis, we have focused on the overall numbers from Steam. Fig.~\ref{fig:1} displays the massive growth of the platform, approximately following a superlinear growth (with power-law exponent equal to $1{.}81$) for the number of games and a linear growth for the number of players as a function of time.
Up to the present moment, there is no apparent sign of saturation, suggesting that this ecosystem still has the potential to grow even further in terms of both games and players. 
From a visual inspection of Fig.~\ref{fig:1}{B}, we can observe the occurrence of well-defined popularity peaks in the second half of the series.
Although we have not calculated specific statistics for these peaks, we have observed that they occur more frequently in the months of January and July, which are typically vacation months. 
It's worth noting that the same conclusions we present throughout this paragraph are applicable when analyzing all the genres (see Figs.~\ref{fig:2},~S1 and~S2).

In an effort to characterize the more general aspects of the game popularity series, we have calculated their average and standard deviations. Fig.~\ref{fig:3}A displays the probability distribution of the number of games as a function of the average popularity fitted by a log-normal. This result is consistent with the probability distribution of the popularity of other kinds of cultural content~\cite{wu2007novelty, yook2020origin}, suggesting the existence of a pattern for the average popularity.
Although it might be intuitive to assume that the distribution of average popularity is Gaussian, especially considering the central limit theorem~\cite{dudley1978central}, we believe that the determining factors for gaining or losing popularity are not additive, but multiplicative.
In other words, we can infer that the collective attention of people is fragmented among the available games and, therefore, this fragmentation process generates a log-normal distribution~\cite{mendes2022popularity, cunha2024complexity}. Interestingly, asymmetric random variables with heavy tails, such as the case of the log-normal distribution, often exhibit rare and extreme events, known as outliers~\cite{downey2023probably}. 
This characteristic is consistent with our findings, where there is a significant difference between a few extremely popular games and the rest. 
However, Fig.~\ref{fig:3}B shows that the dispersion around the average of popularity series is almost always of the same magnitude as the average itself, thus rendering the average an unreliable descriptor for the dynamics of game popularity.


In this uncertain scenario regarding the mean, we have focused on the shape of the game popularity series. 
Knowing that shapes could vary widely (see Fig.~S4), we have adopted a clustering approach.
As described in Sec.~\ref{sec:methods}, the popularity series of each game have been normalized using $z$-scores and then smoothed with a Gaussian filter. 
Despite fixing $\sigma =$ 2 in the Gaussian filter, similar results could be obtained using $\sigma = $ 1 or $\sigma = $ 3. In total, 5 clusters were detected (see Fig.~\ref{fig:4}), which we have named according to their shape pattern: decreasing (cluster 1), hilly (cluster 2), increasing (cluster 3), valley (cluster 4), and bursty (cluster 5). 
The most interesting point in our cluster detection analysis is the connection with other results already documented in the literature.
Firstly, cultural objects undergo a forgetting process, where their popularity exhibits temporal decay~\cite{candia2019universal}. For video games, this result is also supported by highly positively skewed distributions of playtime on Steam~\cite{sifa2015large-scale}.
In this context, we suggest that cluster 1 consists of games whose popularity dynamics are essentially governed by temporal decay. 
Remembering that almost half of the games in our database are in cluster 1, a process of temporal decay can describe the popularity dynamics of almost half of the games. 
Secondly, when evaluating cluster 3, we suggest that its popularity dynamics are largely a result of the preferential attachment mechanism, in which popularity begets popularity~\cite{barabasi1999emergence}.
This cluster represents a small group of games (less than 12\%) with higher popularity averages and continuously increasing behavior over time. 
Along this same line of reasoning, we believe that the pattern of cluster 3 is a strong candidate to describe the popularity of outlier games. At first glance, the detection of outlier games may seem insignificant, but the problem of outliers is one of the oldest in statistics~\cite{hawkins1980identification}.
Thirdly, the late and accelerated growth of popularity in cluster 4 games resembles the dynamics expected of ``sleeping beauties'', first studied with time series of citations of scientific publications. 
In this context, works that remain in a period of hibernation and then experience meteoric popularity are known as ``sleeping beauties''~\cite{ke2015defining}.
Finally, if we intuitively consider that the popularity of games would grow until a saturation value and then decrease to a minimum value (possibly zero)~\cite{ozer2020discovering}, clusters 2 and 5 exhibit the most unexpected patterns.
Cluster 2 represents games with the most unpredictable and irregular popularity series, with oscillations obscuring average trends in some cases. 
We conjecture these oscillations in popularity might be, for example, associated with school breaks and vacations or promoted sales of games at discounted prices.
Cluster 5 presents an entirely unexpected pattern of popularity dynamics, defying our intuition and starting with a certain level of popularity that declines and is gradually recovered after some time.

We also reveal how genres are associated with the patterns detected.
This can be helpful to understanding which games are more likely to achieve increasing or decreasing popularity based on their genres (see Fig.~\ref{fig:5}). In this context, some genres deserve special attention. Indie games are overrepresented in cluster 4, suggesting that many successful Indie games gain popularity belatedly. 
We believe that, being low-budget games, they may struggle to attract attention through large marketing campaigns. 
However, after some time, popularity may grow through word of mouth. Early access games are overrepresented in cluster 3, and our hypothesis for this fact is related to their development process. These games are made available to be played even when incomplete and are updated by developers. In many cases, developers update games for months or even years based on player feedback~\cite{lin2018empirical}. We believe that the result of this development process is games that better please their players. Multiplayer games are overrepresented in clusters 1 and 3, showing that their trajectories tend to be monotonic, increasing or decreasing. 
By definition, these games require people playing together, and we believe this reinforces word of mouth, generating a snowball effect for both the growth or decline of popularity.

Finally, we demonstrate the persistence of shape patterns by comparing the shape of the initial part of a popularity series with the shape of the complete series. Evaluating the first 24 months, we have detected four patterns: decreasing, hilly, increasing, and constant. 
These four patterns give rise to the five patterns already discussed (see Fig.~\ref{fig:sankey_plot}). As a general rule, patterns tend to persist over time, showcasing the consistency of the detected patterns to characterize the dynamics of game popularity. The only pattern that extinguishes over time is the constant pattern, that evolves into hilly (when the variation increases) or bursty (when there is a surge in popularity).

Furthermore, from an economic point of view, game popularity is not a direct measure of financial success.
This is so because monetization depends on the strategy adopted by each developer, with two primary models being games that need to be purchased and free games with paid exclusive content~\cite{olsson2010business}. 
There is also the case of games that monetize in a hybrid way, where it is necessary to buy the game, and there are exclusive contents that can be purchased within the game. Therefore, in addition to contributing to the debate on the dynamics of popularity of cultural objects, our findings may be of some use to the game industry, as understanding the dynamics of popularity can guide developers to better spend on their games.

In conclusion, our research provides novel results about the popularity of video games thus adding to the growing literature of data driven investigations of this cultural product. 
Given the results obtained, we believe our work contributes to the current interest on the existence of universal patterns in popularity dynamics of cultural content. 
Furthermore, our work may inspire further investigations into the characteristics of the shape patterns that define the clusters of popularity series in other contexts.



\clearpage
\newpage
\setcounter{page}{1}
\setcounter{figure}{0}
\makeatletter
\renewcommand{\figurename}{Figure}
\renewcommand{\tablename}{Table}
\renewcommand{\thefigure}{S\@arabic\c@figure}
\renewcommand{\thetable}{S\@arabic\c@table}

\thispagestyle{empty}
\begin{center}
\large{Supplementary Information for}\\
\vskip1pc
\large{\bf Shape patterns in popularity series of video games}\\
\vskip1pc
\normalsize{Leonardo Cunha, Arthur Pessa, Renio Mendes}

\end{center}
\vskip4pc

\begin{table}[!ht]
\centering
\setlength{\tabcolsep}{20pt}
\renewcommand{\arraystretch}{1.5}
\begin{tabular}{| l | c | c |}
\hline
\multicolumn{1}{|c|}{\textbf{Distribution}} &
  \multicolumn{1}{c|}{\textbf{KS-distance}} &
  \multicolumn{1}{c|}{\textbf{KS p-value}} \\ \hline
Log-normal      & 0.0416 & $0.0003$   \\ \hline
Weibull minimum & 0.1051 & $< 0.0001$ \\ \hline
Beta            & 0.1594 & $< 0.0001$ \\ \hline
Double Weibull  & 0.3575 & $< 0.0001$ \\ \hline
Double gamma    & 0.4047 & $< 0.0001$ \\ \hline
Normal          & 0.4701 & $< 0.0001$ \\ \hline
\end{tabular}
\caption{{\bf Kolmogorov-Smirnov Statistical Test for different models of game popularity.} Goodness of fit for different probability distributions modeling the averages of the popularity series. In the first column we have the distributions, in the second column the Kolmogorov-Smirnov distance, and in the third column the p-value resulting from the test. The log-normal distribution has the best fit, with the smallest KS-distance and largest p-value.}
\label{tab_KS_dist_pop}
\end{table}

\begin{figure*}[!ht]
\centering
\includegraphics[width=1\linewidth]{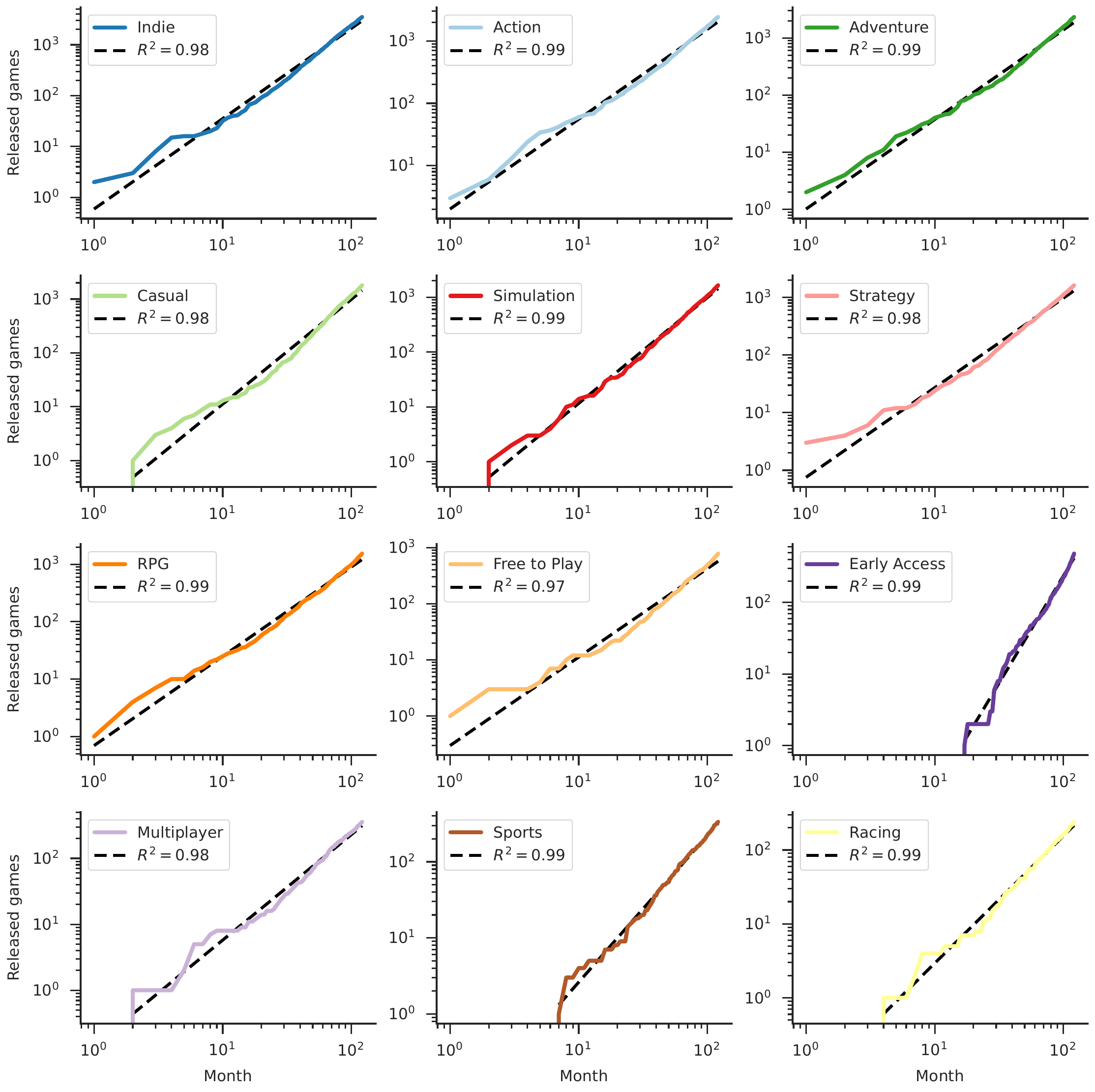}
\caption{{\bf Temporal evolution of the number of games in each genre.} Number of games released for each genre on Steam over the 120 months spanning the period from August 2012 to July 2022. Solid lines represent the data while dashed lines stand for power-law fits.}
\label{fig_s:games_genre_time}
\end{figure*}

\begin{figure*}[!ht]
\centering
\includegraphics[width=1\linewidth]{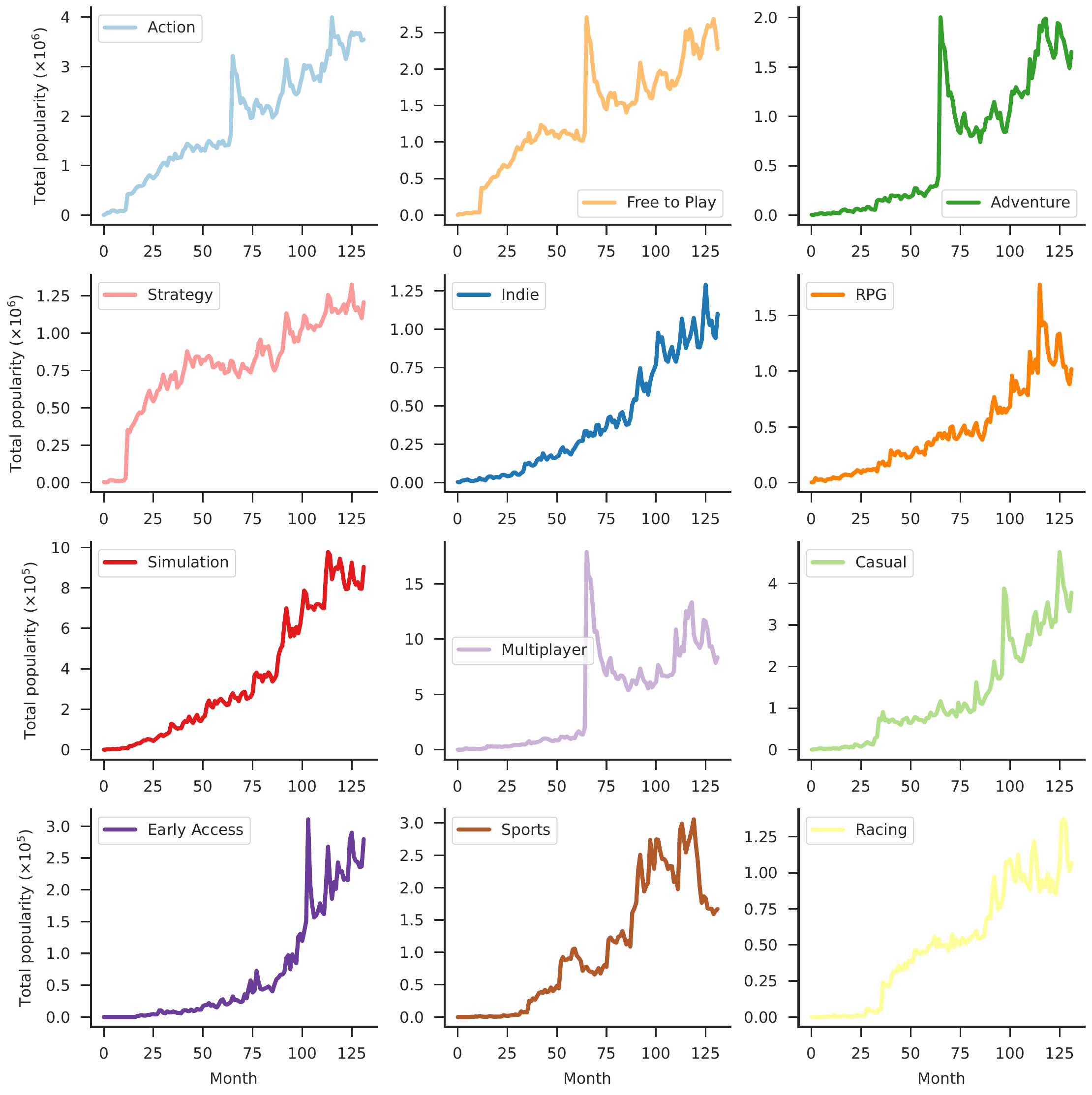}
\caption{{\bf Temporal evolution of the number of players in each genre.} Total popularity in each genre on Steam over the 120 months spanning the period from August 2012 to July 2022.}
\label{fig_s:players_genre_time}
\end{figure*}

\begin{figure*}[!ht]
\centering
\includegraphics[width=1\linewidth]{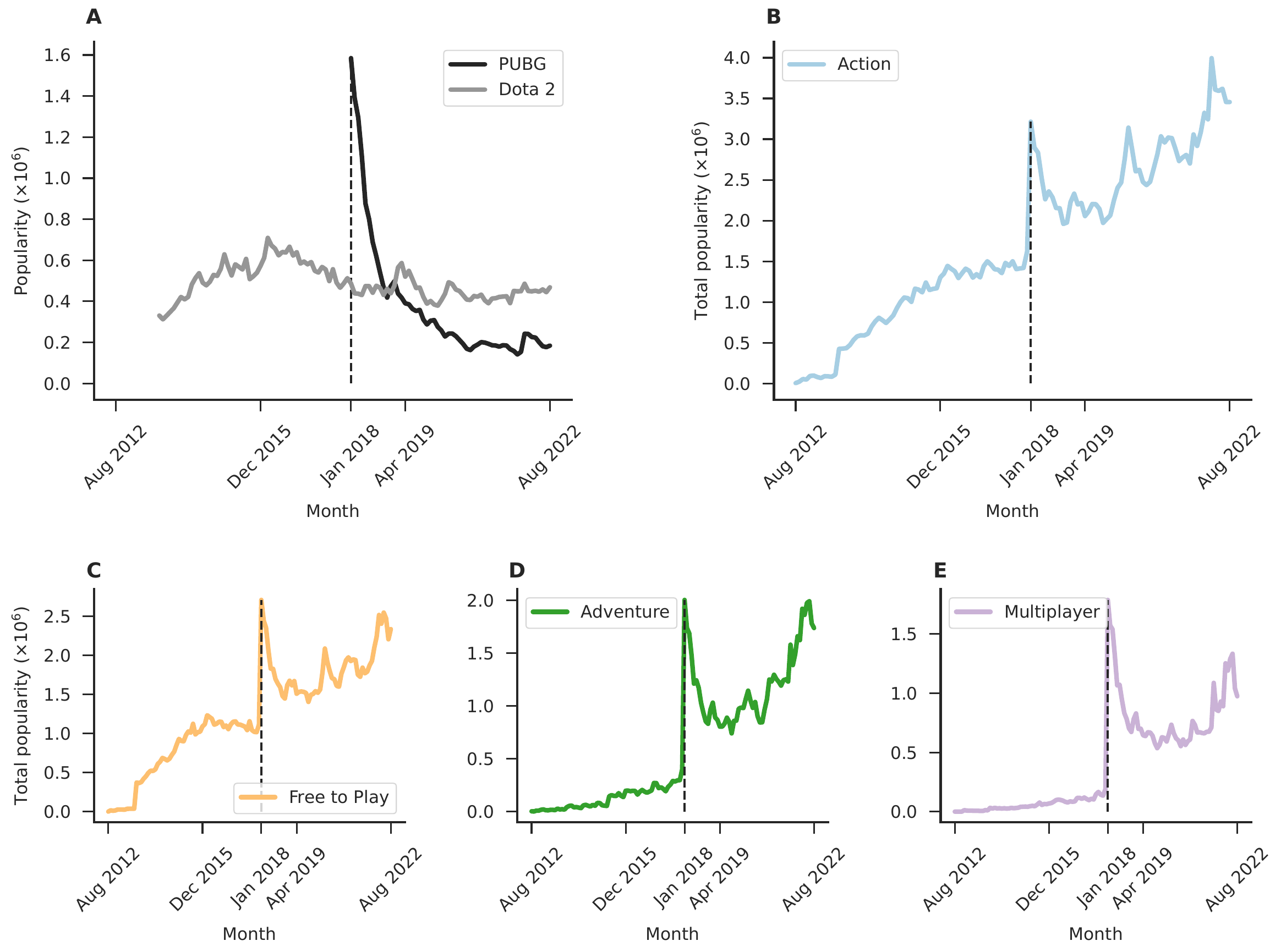}
\caption{{\bf Impact of the release of PUBG.} (\textbf{A}) Popularity series of PUBG (black) and Dota 2 (gray). PUBG was the game with the most players in a single month, reaching its peak number of popularity on the month of its release (January 2018). 
In comparison, Dota 2 was the second game with the most players on Steam in January 2018. 
Panels (\textbf{B})-(\textbf{E}) show the impact of the PUBG release on the popularity of the genres that it is characterized by: Action, Free to Play, Adventure, and Multiplayer.}
\label{fig_s:peak_genres}
\end{figure*}

\begin{figure*}[!ht]
\centering
\includegraphics[width=1\linewidth]{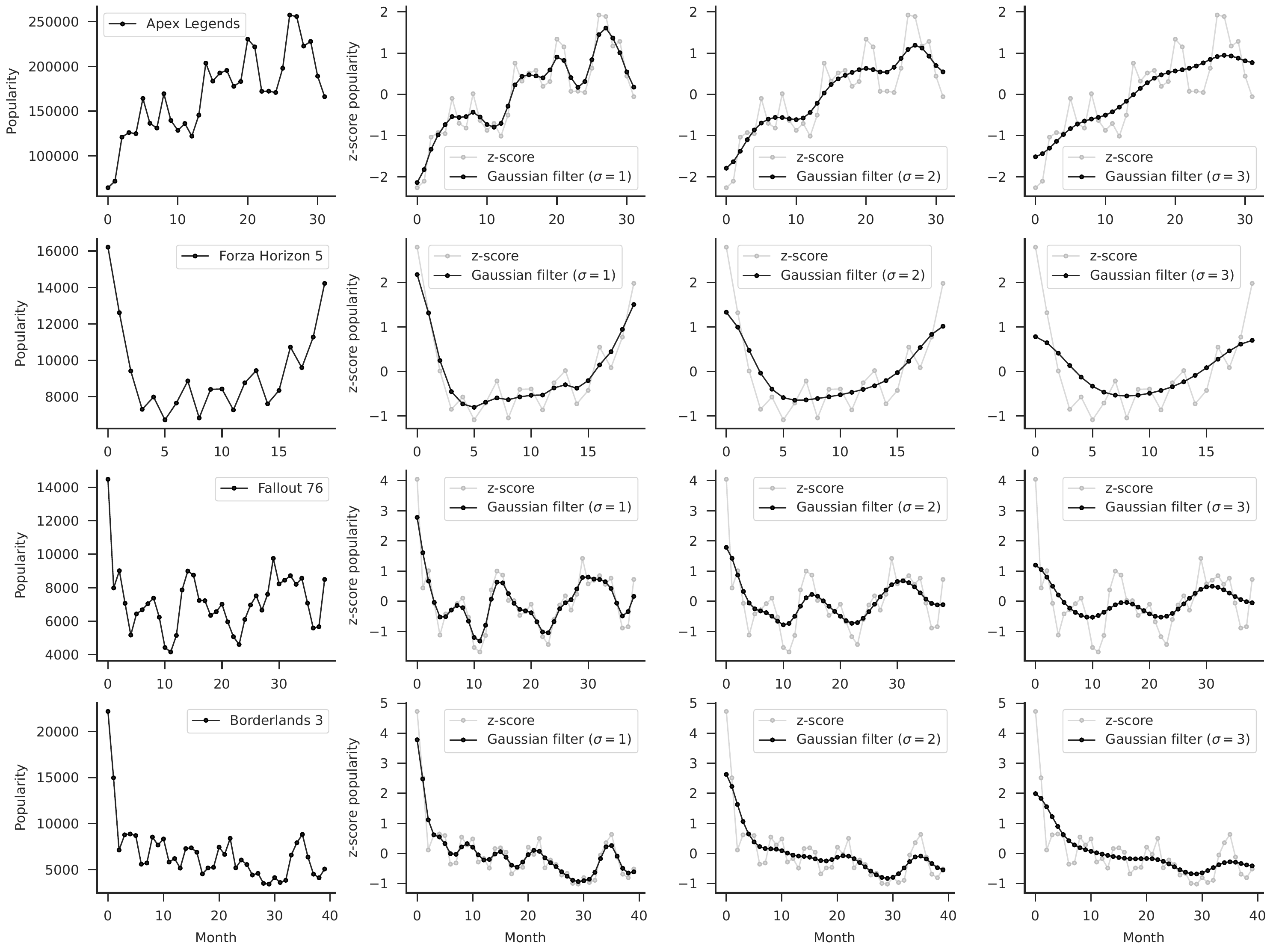}
\caption{{\bf Examples of popularity time series for Steam games and the standardization process of these series.} The first column shows the popularity time series for four games: Apex Legends, Forza Horizon 5, Fallout 76, and Borderlands 3. 
The second, third, and fourth columns show smoothed z-score series using Gaussian filters with $\sigma=$ 1, $\sigma=$ 2, and $\sigma=$ 3, respectively.
In the last three columns, light gray markers represent $z$-score series, while black markers correspond to smoothed $z$-score series.}
\label{fig_s:exemples_series}
\end{figure*}

\begin{figure*}[!ht]
\centering
\includegraphics[width=0.5\linewidth]{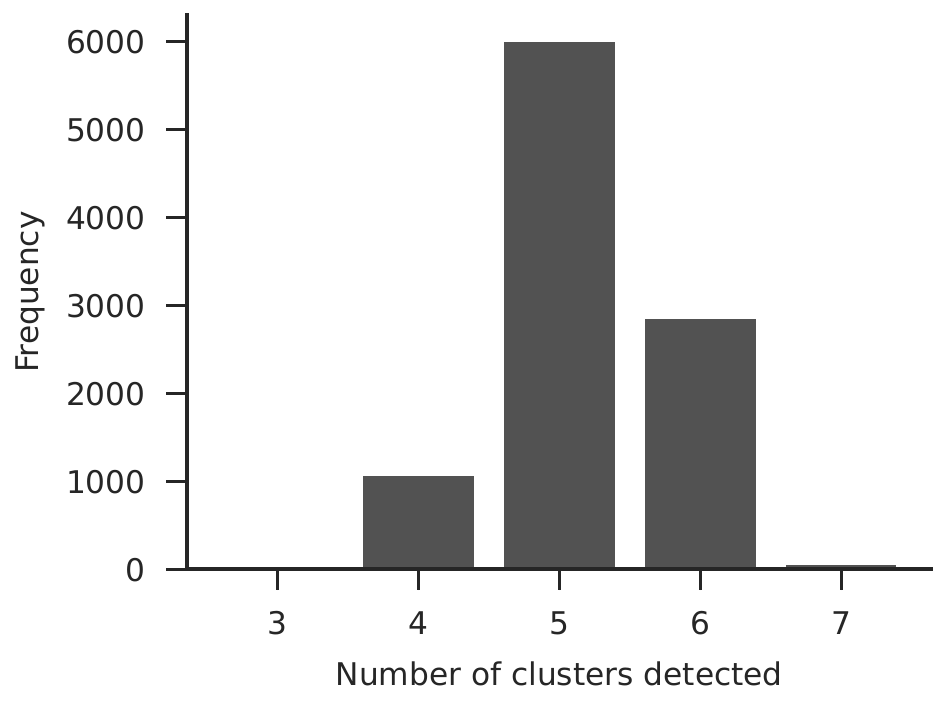}
\caption{{\bf Frequency of the number of clusters detected by the Infomap algorithm.} As the algorithm is sensitive to the input seed, it was executed ten thousand times using different seeds. 
The histogram shows the frequency of the number of clusters detected in all runs. 
The detection of 5 clusters was the mode, and occurred 5,999 times.}
\label{fig_s:n_cluster}
\end{figure*}

\begin{figure*}[!ht]
\centering
\includegraphics[width=0.965\linewidth]{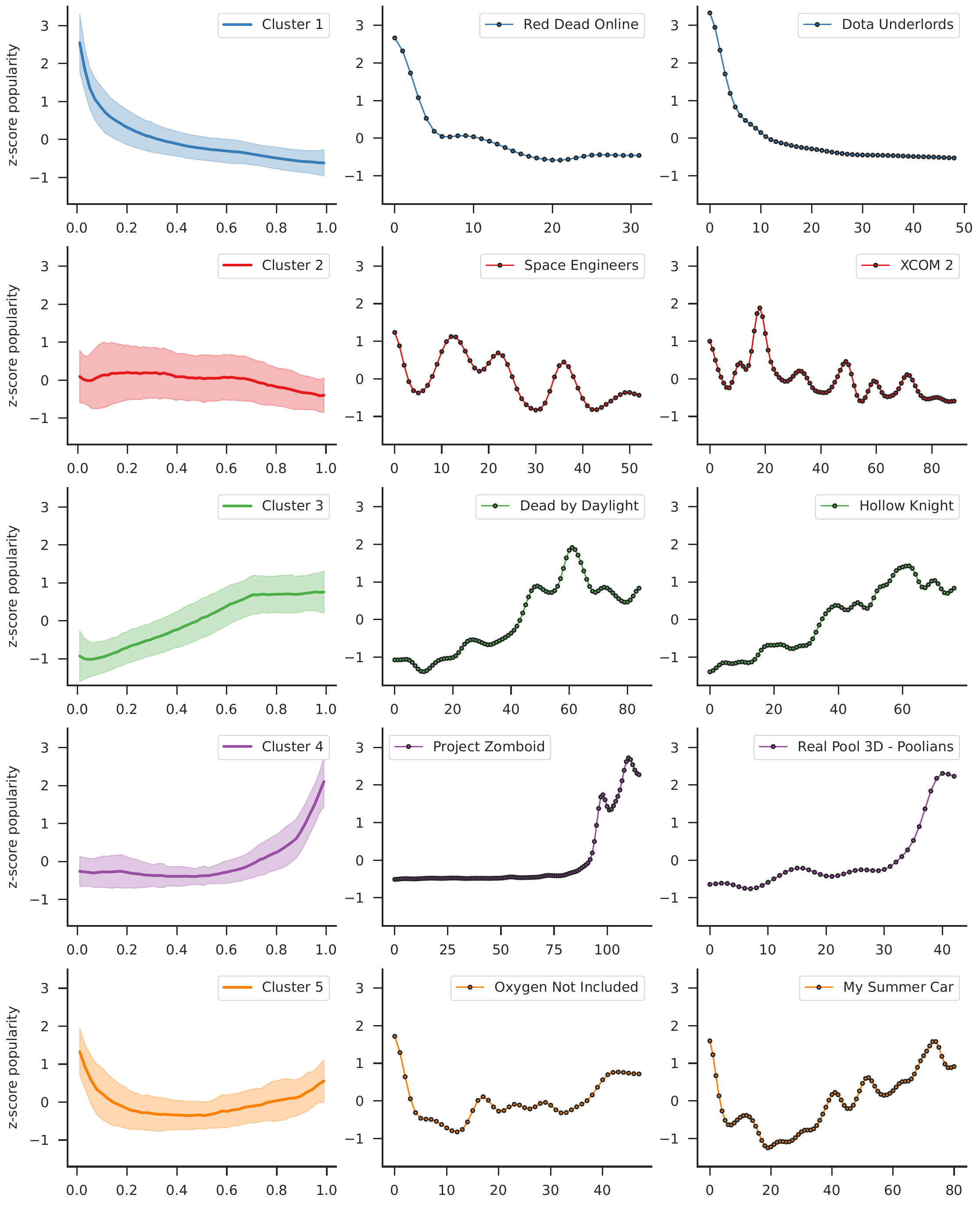}
\caption{{\bf Examples of popularity series for each detected cluster.} The first column displays the popularity pattern for each cluster: cluster 1 (blue), cluster 2 (red), cluster 3 (green), cluster 4 (purple), and cluster 5 (orange). 
The solid lines represent the average of the smoothed $z$-score series for games in each cluster, and the shaded area represents one standard deviation. 
The second and third columns provide examples of smoothed popularity series from games belonging to each cluster.}
\label{fig_s:exemple_clusters}
\end{figure*}

\begin{figure*}[!ht]
\centering
\includegraphics[width=1\linewidth]{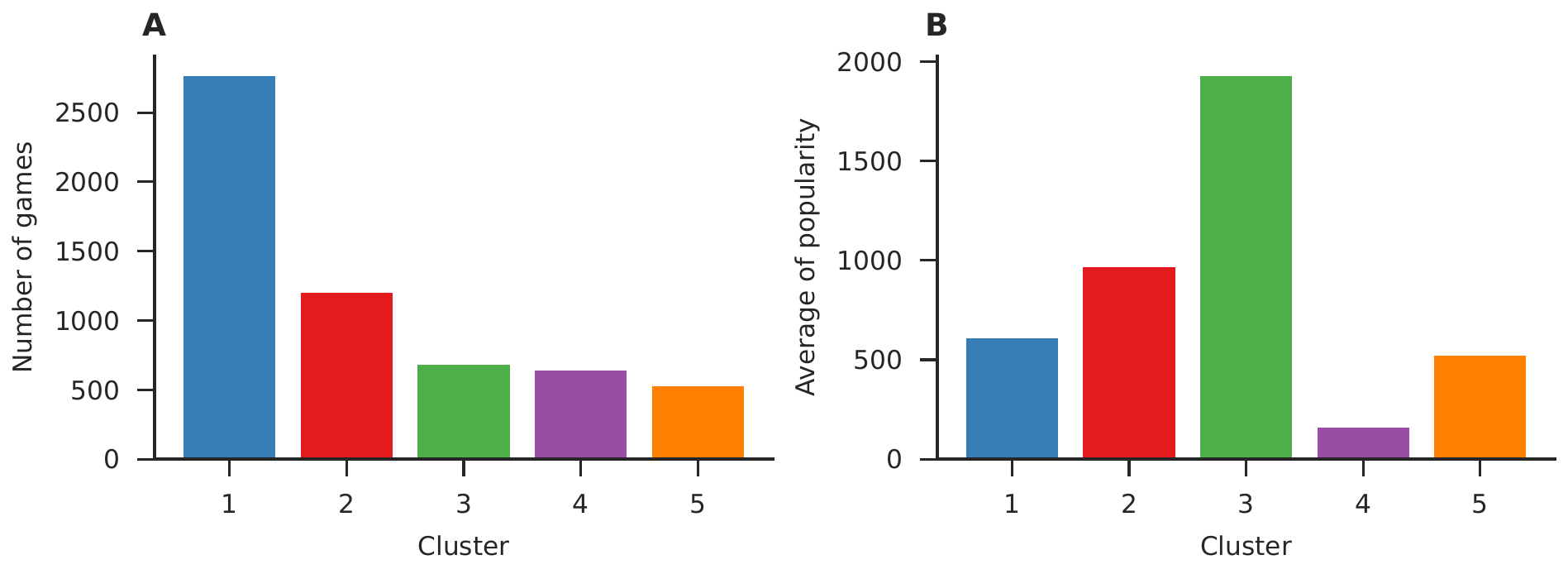}
\caption{{\bf Numbers of games and players associated to the detected clusters.} (\textbf{A}) Number of games in each cluster. (\textbf{B}) Average values of popularity series per game for each cluster.}
\label{fig_s:games_and_players_cluster}
\end{figure*}

\begin{figure*}[!ht]
\centering
\includegraphics[width=0.5\linewidth]{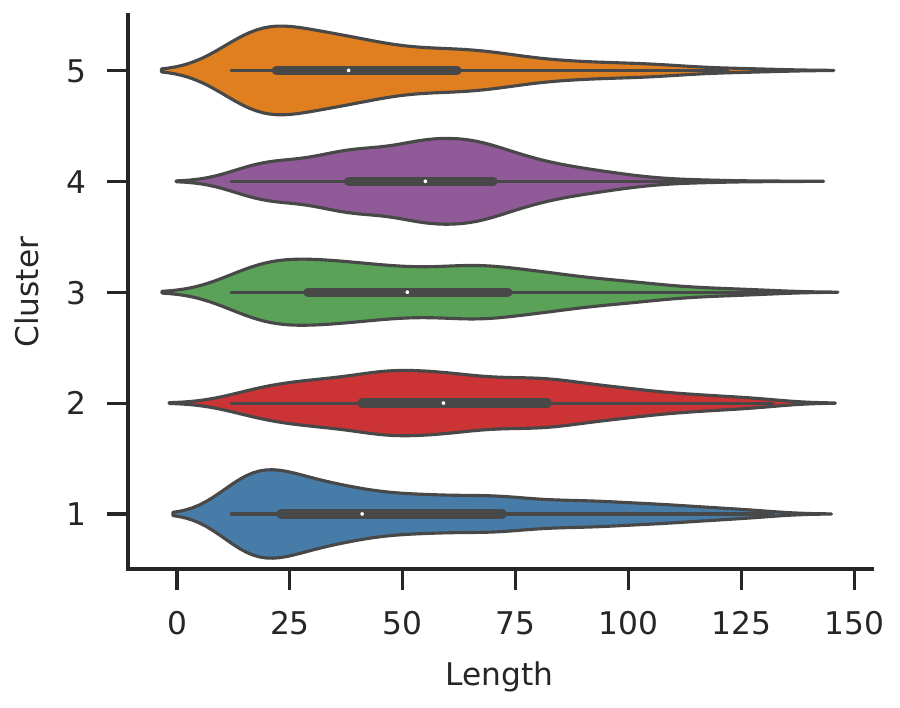}
\caption{{\bf Distribution of series length in each cluster.} Density curves were constructed based on series length data. The $x$-axis marks the series length, the $y$-axis identifies the cluster, and the width of the colored area represents the density of games. For example, clusters 1 and 5 have many games with approximately 25 months of popularity series, while cluster 4 has many games between 50 to 75 months. 
In the density plots, white dots denote the median, black thicker bars represent the interquartile range, and the black thinner lines extend between the lower and upper whiskers.}
\label{fig_s:cluster_length2}
\end{figure*}

\begin{figure*}[!ht]
\centering
\includegraphics[width=0.5\linewidth]{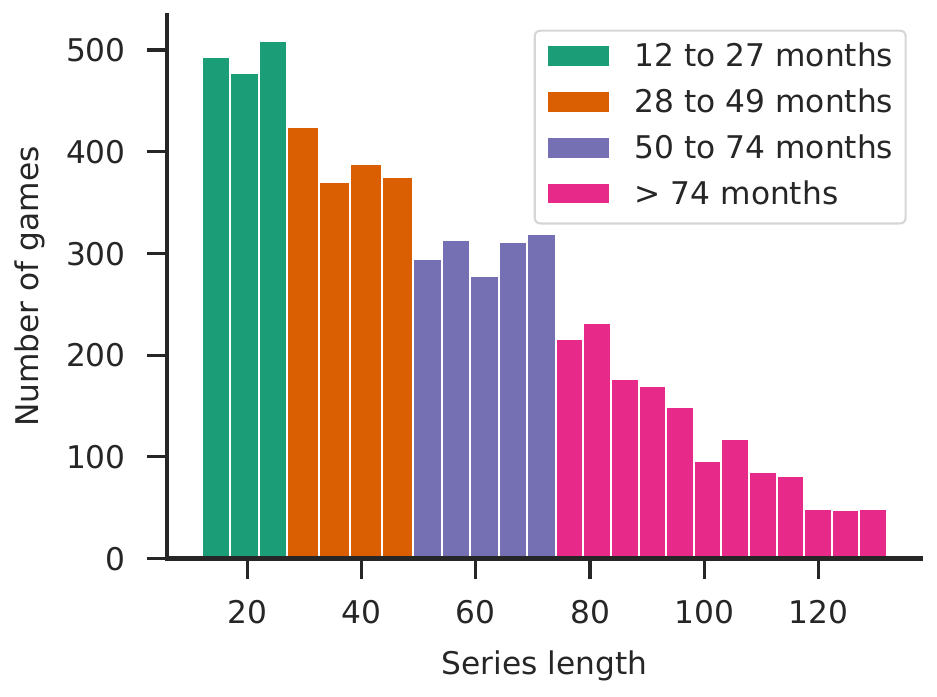}
\caption{{\bf Histogram of game frequency as a function of the length of the popularity series.} Based on this histogram, games were divided into four groups: series between 12 to 27 months long (green), 28 to 49 months (orange), 50 to 74 months (purple), and longer than 74 months (pink).}
\label{fig_s:hist_lenght}
\end{figure*}

\begin{figure*}[!ht]
\centering
\includegraphics[width=0.67\linewidth]{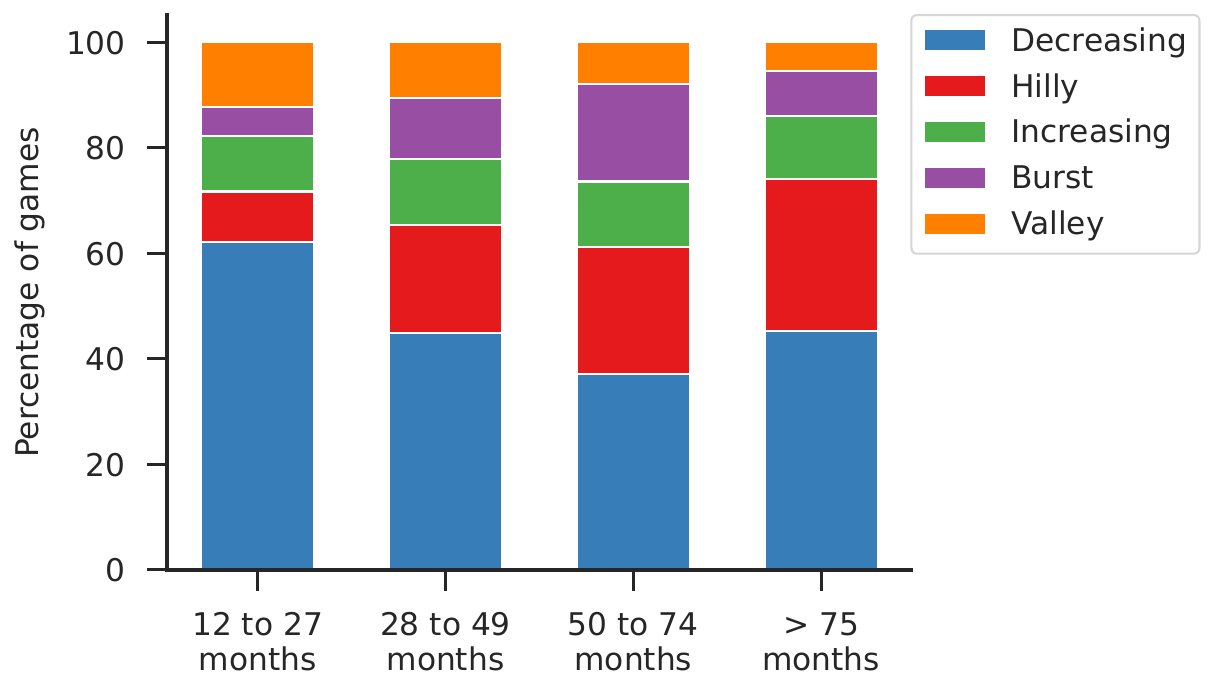}
\caption{{\bf Representation of detected shape patterns in games grouped by series length.}
For each series length group, the percentage of games exhibiting each one of the five detected shape patterns colors a portion of the vertical bar.}
\label{fig_s:cluster_cohort}
\end{figure*}

\begin{figure*}[!ht]
\centering
\includegraphics[width=0.5\linewidth]{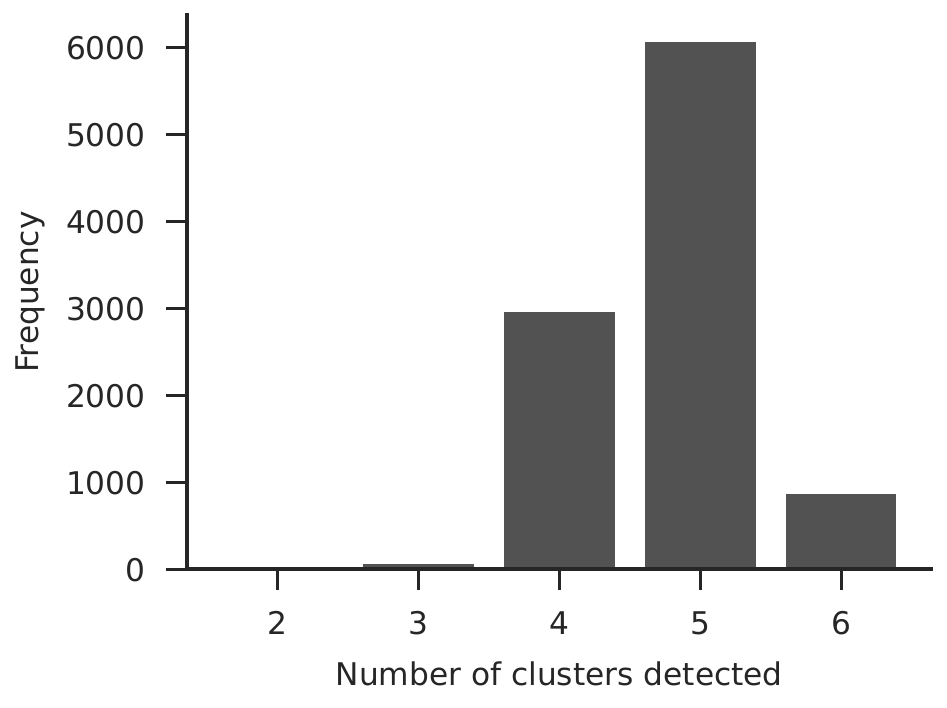}
\caption{{\bf Frequency of the number of clusters detected by the Infomap algorithm for the first 24 months of games with popularity series longer than 74 months.} 
The histogram shows the frequency of the number of clusters detected in ten thousand runs. 
The detection of 5 clusters was the mode, and occurred 6,072 times.}
\label{fig_s:frequency_clusters_24months}
\end{figure*}

\begin{figure*}[!ht]
\centering
\includegraphics[width=1\linewidth]{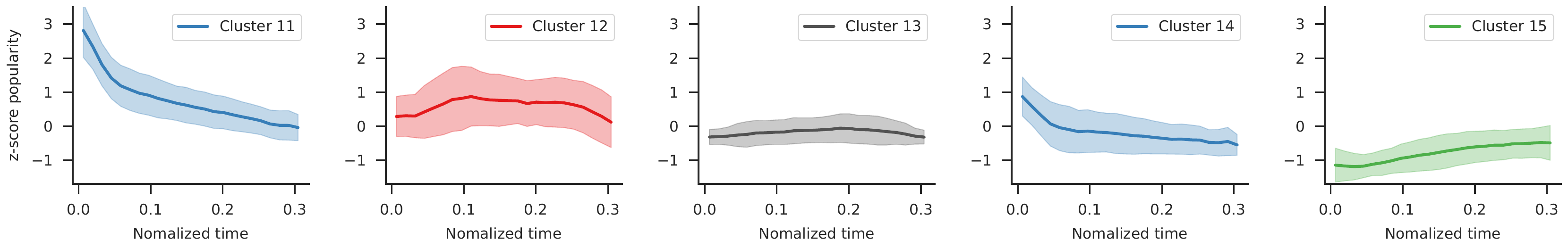}
\caption{{\bf Patterns of popularity series for the first 24 months of games with popularity series longer than 74 months.} The panel displays the average behavior of clusters detected by the best realization of the Infomap algorithm according to the silhouette score. Two clusters exhibited the same pattern: clusters 11 and 14 were displayed a decreasing pattern (blue). 
Another two patterns were once again detected, hilly (cluster 12, red) and increasing (cluster 15, green). 
Cluster 13 (gray) presented a novel constant pattern with relatively little variation around the mean.}
\label{fig_s:clusters_24months}
\end{figure*}

\begin{figure*}[!ht]
\centering
\includegraphics[width=1\linewidth]{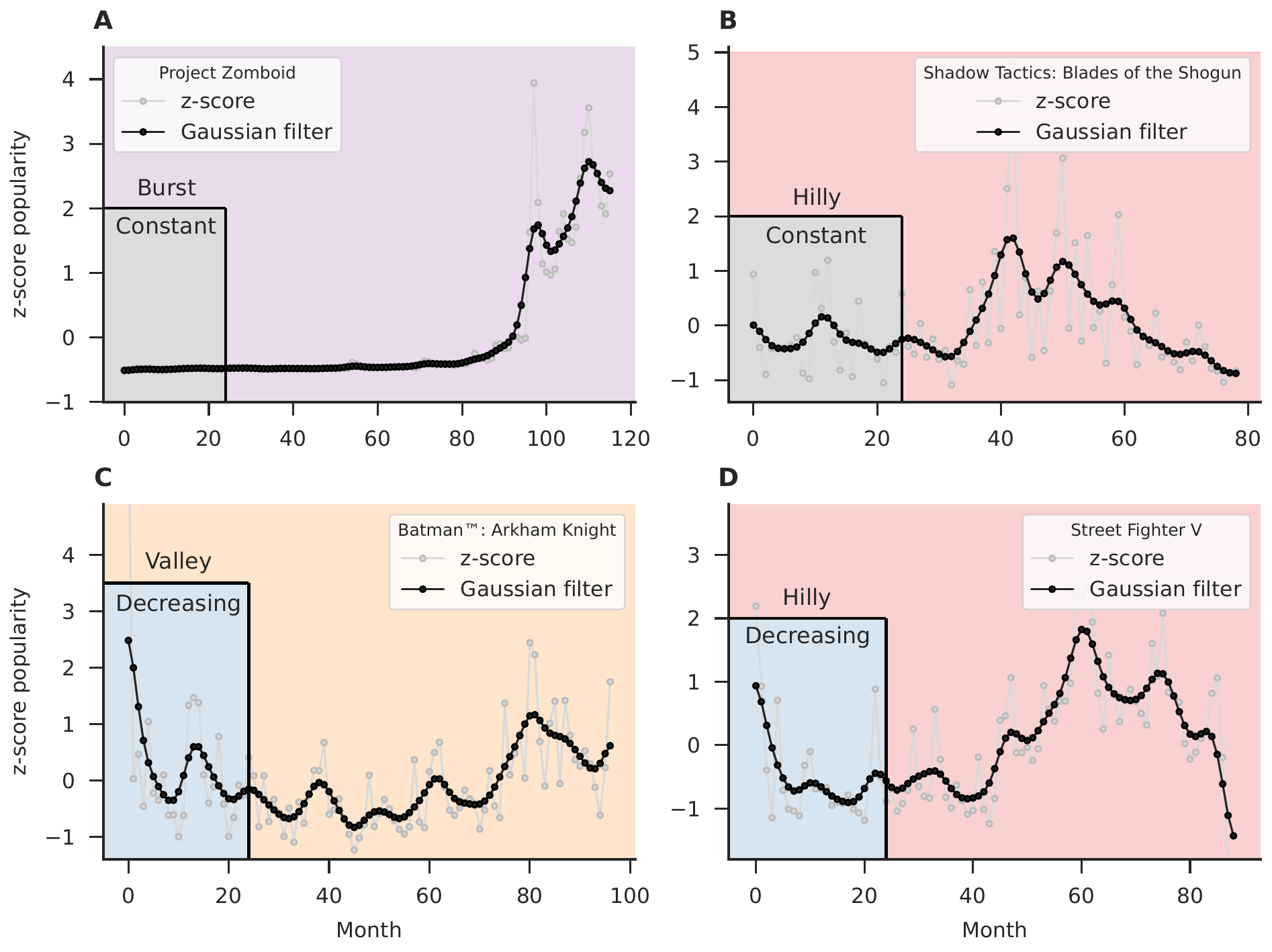}
\caption{{\bf Examples of pattern transition.} Examples of four games (named in within the panels) whose shape of popularity series transition between (\textbf{A}) constant and bursty, (\textbf{B}) constant and hilly, (\textbf{C}) decreasing and valley, and (\textbf{D}) decreasing to hilly.
}
\label{fig_s:exemple_cluster_change}
\end{figure*}

\end{document}